\documentclass[10pt,reprint]{iopart}
\usepackage{graphicx}
\usepackage{psfrag}
\usepackage{color}
\usepackage{amsbsy}
\usepackage{amssymb}
\usepackage{iopams}

\def\@abssec#1{\vspace{.05in}\footnotesize \parindent .2in
{\bf #1. }\ignorespaces}
\newtheorem{theorem}{Theorem}%[section]

 \def \Rm {\mathbb R} \def
\Nm {\mathbb N}  \newcommand{\eps}{\varepsilon}
\newcommand{\dint}{\displaystyle\int}
\newcommand{\dfrac}{\displaystyle\frac}
\newcommand{\pdr}[2]{\dfrac{\partial{#1}}{\partial{#2}}}

\begin{document}

%%%%%%%%%%%%%
%%% TITLE %%%
%%%%%%%%%%%%%
\title{On the reconstruction of diffusions from first-exit time
  distributions}

%%%%%%%%%%%%%%%
%%% AUTHORS %%%
%%%%%%%%%%%%%%%

  \author{Guillaume Bal$^{1}$ and Tom Chou$^{2}$}
\address{$^{1}$Department of Applied Physics and 
        Applied Mathematics, Columbia University, 
        New York NY, 10027, USA; gb2030@columbia.edu}
  \address{$^{2}$Department of Biomathematics, UCLA, Los Angeles, 
      CA, 90095, USA; tomchou@ucla.edu}

%\author{Guillaume Bal \thanks{Department of Applied Physics and 
%        Applied Mathematics, Columbia University, 
%        New York NY, 10027; gb2030@columbia.edu}}
%%%%%%%%%%%%%%%%%%%%%%
%%% BEGIN DOCUMENT %%%
%%%%%%%%%%%%%%%%%%%%%%

%\tableofcontents

\begin{abstract}
  
  This paper explores the reconstruction of drift or diffusion
  coefficients of a scalar stochastic diffusion processes as it starts
  from an initial value and reaches, for the first time, a threshold
  value.  We show that the distribution function derived from repeated
  measurements of the first-exit times can be used to formally
  partially reconstruct the dynamics of the process.  Upon mapping the
  relevant stochastic differential equations (SDE) to the associated
  Sturm-Liouville problem, results from Gelfand and Levitan
  \cite{GelLev-51} can be used to reconstruct the potential of the
  Schr\"{o}dinger equation, which is related to the drift and
  diffusion functionals of the SDE. We show that either the drift or
  the diffusion term of the stochastic equation can be uniquely
  reconstructed, but only if both the drift and diffusion are known in
  at least half of the domain.  No other information can be uniquely
  reconstructed unless additional measurements are provided.
  Applications and implementations of our results are discussed.

\end{abstract}
 
%\begin{AMS}
%\end{AMS}

\renewcommand{\thefootnote}{\fnsymbol{footnote}}
\renewcommand{\thefootnote}{\arabic{footnote}}

%\maketitle

%%%%%%%%%%%%%%%%%%%%%%
%%% BEGINNING TEXT %%%
%%%%%%%%%%%%%%%%%%%%%%

%%%%%%%%%%%%%%%%%%%%%%%%%
\section{Introduction}
\label{sec:intro}
%%%%%%%%%%%%%%%%%%%%%%%%%

%\subsection{Physical Applications -} 

Consider a diffusive stochastic process $X_t$, with an initial
condition $X_0$.  Once $X_t$ reaches a certain threshold value
$X^{*}$, the process is stopped and restarted at $X_0$.  The
distribution of times for $X_t$ to first reach $X^{*}$ across the
ensemble of measurements, the first-exit time distribution, arises in
many physical applications \cite{gardiner,risken}. Mean first-exit
times are readily calculated from explicit forms for the dynamical
rules controlling the process $X_t$. However, in many applications,
the underlying physics is unknown, complex, and/or very difficult to
model. In such instances, the question arises as to whether one can
reconstruct the form of the governing equations or equation parameters
that best fit measured data (such as first-exit time distributions).
In this paper, we analyse a general scalar diffusive process to
formally determine how much of the drift and diffusion functions can
be determined from first-exit time data.

Many first passage time problems can be recast as inverse problems.
We describe two physical realizations that motivate this work:
transmembrane voltage spike frequencies in neurons
\cite{gerstner,Tuckwell89}, and first passage times over a molecular
barrier in the rupture of chemical bonds \cite{gardiner,kramers}.  The
transmembrane voltage of a nerve cell is routinely measured and
typically exhibits repeated, sharp spikes.  A class of reduced models
for the transmembrane voltage $V$ is described by

\begin{equation}
\label{eq:voltage}
{d V_t \over dt} = I(V_t) + g(V_t) \eta_t,
\end{equation}

\noindent 
where $I(V_t)$ and $g(V_t)$ are typically polynomial functions of $V$,
and $\eta_t$ is a noise arising from interactions with the many other
connected neurons. This Langevin-type equation is to be appropriately
interpreted with Stratonovich calculus \cite{VK}. An additional,
instantaneous nonlinearity that gives rise to a spike and voltage
reset is implicitly included by imposing a threshold voltage $V^{*}$.
This nonlinearity arises ultimately from the nonlinear dynamics of ion
channels that span the cell membrane \cite{gerstner}.  Upon reaching
$V^{*}$, the system instantaneously spikes and resets to the value
$V_{0}$. Various forms of $I(V_t)$ and $g(V_t)$ have found wide use.
For example, when the noise $\eta_t$ arising from other connected
nerve cells changes the nerve cell's membrane conductivity, $g(V_t)
\propto V_t + \mbox{const.}$ Both linear (linear integrate-and-fire)
and quadratic (quadratic integrate-and-fire) forms are used to model
$I(V_t)$ \cite{gerstner,nykamp}, although it is often unclear which
model of $I(V_t)$ is most appropriate \cite{nykamp}.  Instead of
solving for the spike times for presupposed $I(V_t)$ and $g(V_t)$
\cite{Tuckwell80}, our goal is to infer as much as possible about the
nonlinear current-voltage relationship $I(V_t)$ and coupling to noisy
inputs $g(V_t)$, given a measured spike time distribution
\cite{Tuckwell78}.

Another application where first exit times can be used to reconstruct
models is in the single molecule measurements of macromolecular
detachment, or bond rupturing. An example of this rapidly developing
field is the accelerated rupturing of bonds upon application of a
dynamic load between two macromolecules, such as the ligand-receptor
complex biotin-streptavidin \cite{pull1,pull2}.  The mean times to
bond breaking have been measured \cite{pull1} and modelled
\cite{pull2}. This ``Kramers problem'' \cite{gardiner,kramers,risken},
can also be studied as an inverse problem.  Bond rupture can be
modelled as the irreversible passage of the relevant molecular
coordinate $X$ over a barrier from a metastable state.  Thermal
bombardments agitate both the molecular distance coordinate directly
and the interaction potential.  In the overdamped, or Markovian limit,
the distance coordinate obeys the Langevin equation

\begin{equation}
{d X_t \over dt} = - U'(X_t) + g(X_t) \eta_t,
\end{equation}

\noindent 
where $U(X_t)$ is the effective interaction potential of the rupturing
bond, and $g(X_t)$ represents a force that arises from a fluctuating
potential driven by the white diffusion noise $\eta_t$ (again, the
appropriate convention is the Stratonovich difference rule). The {\it
  forward} problem of bond rupture, or detachment, using simple fixed
potentials $U(X_t)$ \cite{gardiner,kramers}, and with explicit forms
for fluctuating potentials \cite{doering}, have been well-studied.
 
\begin{figure}
    \begin{center}
      \includegraphics[height=2.0in]{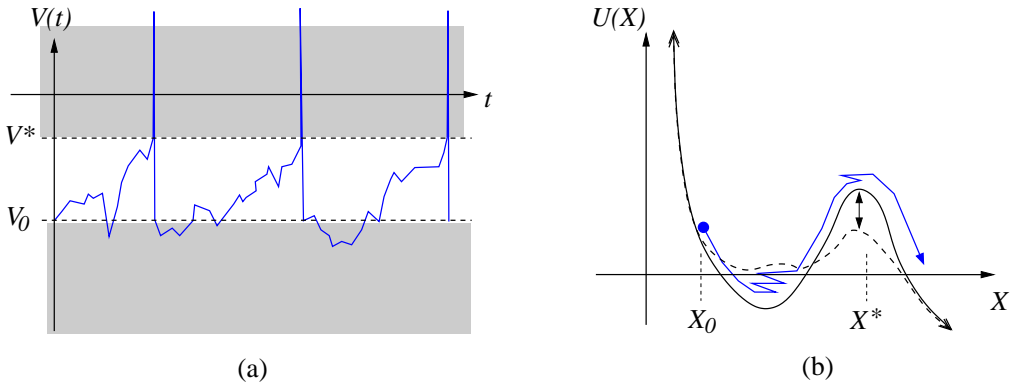}
    \end{center}
    \caption{(a) The transmembrane potential is instantaneously reset to 
      $V_{0}$ as soon as it crosses $V^{*}$. We demonstrate that the
      functional form of the voltage dynamics can be be determined
      only between $V_{0}$ and $V^{*}$ (the unshaded region).  (b)
      Passage over a metastable barrier.  The first-passage,
      bond-breaking times can be used to construct the potential
      $U(X)$ and and how it fluctuates (dashed curve).}
    \label{Fig1}
\end{figure}

Both problems discussed above (see Fig. \ref{Fig1}), as well as many
others, can be modeled using a scalar stochastic process where the
first-exit time distribution obeys the backward Kolmogorov equation.
We show in the following sections, that the measured first-exit time
distribution (given a fixed initial condition) along with knowledge of
the drift and diffusion parameters in part of the domain, is
sufficient to uniquely determine one of them in the entire domain
provided the other one is known. Our analysis gives a formal procedure
for reconstructing drift or diffusion of the stochastic process.  It
is based on transforming the equation into a Schr\"odinger type
equation and on analytical (more precisely meromorphic) continuation
of measured data to obtain the spectral data of the Schr\"odinger
equation.  Although uniqueness is demonstrated in the aforementioned
framework, the necessary analytical continuation in the reconstruction
shows that the inverse problem is {\em severely ill-posed}
\cite{engl}.

%%%%%%%%%%%%%%%%%%%%%%%%%%%%%%%%%%%%%%%%%%%%%%
\section{Stochastic Equations and Main Result}
\label{sec:stoch}
%%%%%%%%%%%%%%%%%%%%%%%%%%%%%%%%%%%%%%%%%%%%%%

Consider the general stochastic process

\begin{equation}
  \label{eq:process}
  \begin{array}{l}
  dX_t = \mu(X_t) dt + \sigma(X_t) d W_t,
  \end{array}
\end{equation}

\noindent where $dW_t= \eta_tdt$ is the Wiener measure, and 
the It\^o interpretation is used \cite{VK,oksendal}. The drift term,
$\mu(X)$, is assumed to have sub-linear growth at infinity, and is of
class $C^1(\Rm)$. We also assume that the variance, $\sigma(X)$, is
uniformly bounded from above and below by positive constants, and is
of class $C^2(\Rm)$.  These assumptions are sufficient to ensure that
the process (\ref{eq:process}) admits a unique solution
\cite{oksendal}.

In the application to spiking action potentials in neurons, $X_t
\equiv V_t$ is the transmembrane voltage at time $t$. The potential
immediately after reset is denoted $X_{0}\equiv V_{0}$. The
corresponding drift and diffusion terms are defined as $\mu(X) \equiv
I(X) + {1\over 2}g(X)g'(X)$ and $\sigma(X) \equiv g(X)$, respectively,
since (\ref{eq:process}) is in the It\^o form and (\ref{eq:voltage})
should be interpreted in the Stratonovich sense \cite{oksendal}.  For
a more detailed derivation of the the diffusion process in the context
of action potential modeling, we refer the reader to number of
standard works \cite{gerstner,johnston,Lange,Tuckwell89}.  In the
application to chemical bond rupturing, $X$ denotes the reaction
coordinate along an effective molecular energy landscape (cf. Fig.
1(b)), $X_{0}$ is an initial bond displacement, and we identify the
drift in (\ref{eq:process}) with $-U'(X) + {1\over 2}g(X)g'(X)$.

Let us now consider the interval $\Delta =(X_{-},X^{*})$ and the first
exit time $t^*$ of the process $X_t$; {\it i.e.}, $t^*$ is the first
time of exit of $X_t$ out of $(X_-,X^*)$
\cite{breiman,Lange,oksendal}.  First exit time measurements can then
be used to reconstruct properties of the diffusion process.  For
instance, it is known that knowledge of $P_X(X_{t^*}=X^*)$ (the
probability that $X_t$ exits $\Delta$ at $X^*$ conditional on it
starting at $X\in \Delta$) and ${\bf E}_X[t^*]$ (the expected time it
takes for $X_t$ to exit $\Delta$ starting at $X$), uniquely determines
$\mu$ and $\sigma$.  We refer to Breiman \cite{breiman} for details on
the so-called {\em natural time scale} and {\em speed measure} of
diffusion processes.

In the applications presented here, the situation is different: We
have access to the full distribution of the first-exit times, but {\em
  only} for processes starting at $X = X_0$ (the reset potential or
the initial relative molecular coordinate).  The distribution of the
exit times can be conveniently written as the solution to a partial
differential equation (the backward Kolmogorov
equation)\cite{gardiner,GS72,Lange,risken,Tuckwell76,Tuckwell89}.  Let
us denote by $w(X;\lambda)$ the Laplace transform (from the time
variable $t$ to the dual variable $\lambda$) of the probability
density $\tilde w(X;t)$ of the first-exit times $t$ from $X \in
\Delta$ of the diffusion process starting at $X=X_{0}$.  For
simplicity, we will assume reflecting boundary conditions for the
diffusion at $X=X_{-}$, and consider termination of the process only
at $X^{*}$. Following \cite{gardiner,GS72,Lange,Tuckwell76}, we find
that $w(X;\lambda)$ solves the following backward Kolmogorov equation,

\begin{equation}
  \label{eq:wx0}
  \begin{array}{l}
\fl \displaystyle   \frac{\sigma^2(X)}{2}w''(X;\lambda)+\mu(X)w'(X;\lambda)
=\lambda w(X;\lambda),
    \,\,\, X_{-} < X < X^{*},\,\,\, \lambda>0, \\
  w(X^{*};\lambda)=1, \quad \displaystyle 
   {\partial w(X_{-};\lambda) \over \partial X} = 0,
 \end{array} 
\end{equation}

\noindent 
where $\partial w(X_{-};\lambda)/\partial X = 0$ corresponds to
vanishing probability flux at $X=X_{-}$. Although we restrict our
analysis to a reflecting lower bound, depending on the physical
application, other boundary conditions can be imposed at an arbitrary
lower bound.  The boundary condition $w(X^{*};\lambda)=1$ reflects the
fact that $\tilde w(X_{0}=X^{*},t) = \delta(t)$.  Since $\tilde
w(X_{0};t)$ is directly measured, $w(X_{0};\lambda)$ is exactly
determined for all $\lambda>0$.  Using an affine change of variables,
$(X-X_{-})/(X^{*}-X_{-}) \rightarrow x$, we rescale the domain to
$x\in [0,1]$ and recast Eq. (\ref{eq:wx0}) as

\begin{equation}
  \label{eq:wx}
  \begin{array}{l}
  \displaystyle  \frac{\sigma^2(x)}{2} w''(x;\lambda)+\mu(x) 
   w'(x;\lambda)=\lambda w(x;\lambda),
   \quad 0 < x < 1,\quad \lambda>0,\\
  w(1;\lambda)=1, \quad\displaystyle{\partial w(0;\lambda)\over\partial x}=0,
 \end{array} 
\end{equation}
where $x$ is now the initial position of the process. We still use the
notation $\sigma$ and $\mu$ for the diffusion and drift terms in the
rescaled variables. For example, in the neuron spiking problem, the
voltage threshold is at $1$, the inhibitory threshold (the lower bound 
for the voltage) is at $0$, and the reset voltage that
restarts the process is at $x=x_{0}$. This problem admits a unique solution
provided $\sigma$ remains uniformly bounded from above and below and
$\mu$ remains bounded.

In many instances $\mu(x)$ and $\sigma(x)$ are known from a
well-accepted model, and the function $w(x;\lambda)$ can be computed
directly.  Now consider the case where $\mu(x)$ and $\sigma(x)$ are
unknown, but $w(x=x_{0};\lambda)$ is {\it measured} at a single
initial position $x_{0}= (X_{0}-X_{-})/(X^{*}-X_{-})$.  What can one
reconstruct (with respect to $\mu(x)$ and $\sigma(x)$) from a
knowledge of $w(x_{0};\lambda)$ for all $\lambda>0$?  In answering
this question, we arrive at the following:

Let $0<x_b<1$ be the point defined in (\ref{BCOND}) below and let
$\Lambda\subset(0,1)$ be a countable set of Lebesgue measure $0$
in $(0,1)$, defined in (\ref{eq:countset}) below. We obtained the 
following related results:
\begin{theorem}
  \label{thm:1}
  Assume that $\mu(x)$ and $\sigma(x)$ are known on $(0,x_b)$ and that
  the measurement point $x_0\in (0,1)\backslash\Lambda$.  If, in
  addition, either $\mu(x)$, $\sigma(x)$, or $\mu(x)/\sigma(x)$ is
  known entirely on $[0,1]$, then both $\mu(x)$ and $\sigma(x)$ are uniquely
  determined on $[0,1]$ by measurement of $w(x_{0};\lambda)$, for all
  $\lambda>0$.
\end{theorem}
\begin{theorem}
  \label{thm:2}
  Assume that $\mu(x)$ and $\sigma(x)$ are known on $(0,x_0)$ and that
  the measurement point $x_0\geq x_b$.  If, in addition, either
  $\mu(x)$, $\sigma(x)$, or $\mu(x)/\sigma(x)$ is known on $[0,1]$,
  then both $\mu(x)$ and $\sigma(x)$ are uniquely determined on
  $[0,1]$ by measurement of $w(x_{0};\lambda)$, for all $\lambda>0$.
\end{theorem}

Both theorems say that if either $\mu$ or $\sigma$ is known on
$[0,1]$, while the other is known only on half of the domain, we can
reconstruct this other function also on the entire domain.  Since
$x_{b}$ is implicitly defined by $\sigma$ (cf. Eq. \ref{BCOND}), it is
an {\it a priori} constraint on the size of the domain that one can
reconstruct from the measurements $w(x_{0};\lambda)$. In {Theorem 1},
the measurements are made at an almost arbitrary point $x_0\in(0,1)$.
We have to remove a countable set of points, which is also defined
implicitly as it depends on $\sigma$ and $\mu$.  This constraint is no
longer necessary in {Theorem 2}, when $x_0\in(x_b,1)$.

The conclusions of both results hold true if the domains $(0,x_b)$ and
$(0,x_0)$ of {\it a priori} knowledge of the coefficients are replaced
by $(x_b,1)$ and $(x_0,1)$, respectively with $x_0\leq x_b$ in the
latter case. We do not explicitly consider these cases. The main steps
of the proof are the following. First we use a transformation similar
to the Liouville transformation (used to obtain the canonical form of
Sturm-Liouville problems \cite{CCPR-SIAM97}) to map Eq. (\ref{eq:wx})
into a Schr\"odinger equation (Section \ref{sec:sch}). Next, we show
that a measurement of $w(x_{0};\lambda)$, for all $\lambda>0$, is
sufficient to obtain a single eigenvalue spectrum of the
Sturm-Liouville problem (when $x_0\not\in\Lambda$ in the setting of
{Theorem 1}).  This allows us to use results from Gelfand and Levitan
\cite{GelLev-51} to verify uniqueness of the reconstruction (Section
\ref{sec:rec}) and explicitly extract $\mu(x)$ or $\sigma(x)$ (Section
\ref{sec:recfinal}). The above theorems require to know the
coefficients on half of the domain before one of them can be
reconstructed on the other half. An explicit asymptotic expansion
presented in section \ref{sec:recfinal} shows that this a priori
knowledge is indeed necessary to obtain a unique reconstruction.  In
section \ref{sec:multiple} we present a result showing that the
reconstruction of one of the coefficients $\mu(x)$ or $\sigma(x)$ is
unique over the whole domain $(0,1)$ provided that they are known on
an arbitrarily small domain $(0,\eta)$ and provided that a sequence of
measurements $w(x_m;\lambda)$ is available for $\lambda>0$ and $x_m$,
$m\in\Nm$, judiciously chosen. In this result, a minimum of $M$
measurements is required to reconstruct $\mu$ or $\sigma$ provided
that they are known {\it a priori} on an interval close to $x=0$ of
size $2^{-M}$. Finally section \ref{sec:impl} considers possible
implementations of the method.

%%%%%%%%%%%%%%%%%%%%%%%%%%%%%%%%%%%%%%%
\section{Mapping to Schr\"odinger form}
\label{sec:sch}
%%%%%%%%%%%%%%%%%%%%%%%%%%%%%%%%%%%%%%%

To recast Eq. (\ref{eq:wx}) as a Schr\"odinger equation, we effect the
following change of variables
\begin{equation}
  w(x) = (fu)\circ h(x).
\end{equation}

\noindent Thus,

\begin{equation}
  w'(x) \Rightarrow h' (fu)'\circ h,\qquad w''(x) \Rightarrow h'' 
(fu)'\circ h + (h')^2 (fu)''\circ h, 
\end{equation}
and Eq. (\ref{eq:wx}) becomes
\begin{equation}
  \label{eq:wx2}
 \Big[\frac{\sigma^2}{2} (h')^2 \Big] (fu)''\circ h
   + \Big[ \frac{\sigma^2}{2} h'' + \mu h'\Big] (fu)' \circ h 
   =\lambda (fu)\circ h.
\end{equation}

\noindent To normalise the second-order term, 
we impose the constraint

\begin{equation}
  \label{eq:h}
  \begin{array}{l}
  \displaystyle h'(x) = \frac{\sqrt2}{\sigma(x)} ,\qquad 0 < x <1, \\
 h(0)=0.
  \end{array}
\end{equation}
Since $\sigma$ is bounded from above and below by a positive constant,
$x\to h(x)$ is a diffeomorphism. Upon introducing the change of
variables
\begin{equation}
  \label{eq:yofx}
  y = h(x),
\end{equation}
the points $x=0$, $x=x_0$, and $x=x_b$ before the change of variables
become $y=h(0)=0$, $y_{0}\equiv h(x_{0})$, and $y_{b} \equiv h(x_b)$,
respectively.  If we define a new drift
\begin{equation}
  \label{eq:nu}
  \nu(y) = \Big(\frac{\sigma^2}{2}h''+\mu h'\Big)(h^{-1}(y)), 
   \qquad 0 \leq y \leq y^{*}\equiv h(1),
\end{equation}
Eq.  (\ref{eq:wx2}) can be succinctly expressed as
\begin{equation}
  (fu)''(y) + \nu(y)(fu)'(y) =\lambda (fu)(y) .
\end{equation}
To remove the drift term, we force $f$ to be the solution to
\begin{equation}
  \label{eq:f}
 \begin{array}{l}
  2f'(y) + \nu(y) f(y) =0, \qquad 0 < y < y^{*}, \\  f(y_{0})=1.
  \end{array}
\end{equation}
Since $f(y)>0$ for $y=y_{0}$, $f(y)>0$ for all $y \in [0,y^*]$.
We finally obtain
\begin{equation}
  \label{eq:schforu}
  \begin{array}{l}
   u''(y) - q(y) u(y) =\lambda u(y), \qquad 0 < y < y^{*}\\
  u(y^{*})=\dfrac{1}{f(y^{*})}>0, \quad 
  \displaystyle {\partial u(0;\lambda) \over \partial y} = 0, 
  \end{array}
\end{equation}
where the potential is defined by
\begin{equation}
  \label{eq:q}
  q(y) = -\frac{\nu f'+f''}{f}(y).
\end{equation}

By assumption, $\sigma(x)$ and $\mu(x)$ are known for $0\leq x\leq b$.
This implies that $\nu(y)$ and $f(y)$, and hence $q(y)$, are known for
$0 \leq y\leq y_{b}\equiv h(x_b)$, although $y^{*}\equiv h(1)$ itself is
generally {\em not} known (unless $\sigma(x)$ is also known on
$[0,1]$).  This undetermined domain size is found in the following
section.

%%%%%%%%%%%%%%%%%%%%%%%%%
\section{Uniqueness of the potential reconstruction}
\label{sec:rec}
%%%%%%%%%%%%%%%%%%%%%%%%%
We now consider the reconstruction of the potential $q(y)$ in the
equation
\begin{equation}
  \label{eq:sch}
  \begin{array}{l}
  u''(y;\lambda) - q(y) u(y;\lambda)=\lambda u(y;\lambda), \qquad 0 <y 
  < y^{*}\equiv h(1) \\
  u(y^{*})=\dfrac{1}{f(y^{*})}>0, \quad 
   \displaystyle {\partial u(0;\lambda)\over \partial y}=0.
%\quad u(h(X_{-}) = 1/f(h(X_{-}) \mbox{ or } u'+{f'\over f}u(h(X_{-}))=0,
%\quad |u(x;\lambda)|\leq C.
  \end{array}
\end{equation}
The assumption is that $q(y)$ is known on $[0,y_{b}]$ as described
earlier. However $y^{*}=h(1)$ depends on $\sigma(x)$ throughout the
domain and is unknown, as is $q(y)$ on $(0,y^{*})$.  From
measurements, we have $u(y_{0};\lambda)$ for all $\lambda>0$ since
$w(x_{0};\lambda)=u(y_{0};\lambda)$ and $f(y_{0})=1$.

Uniqueness of the reconstruction of $q(y)$ on $(y_b,y^*)$ is based on
the fact that $\lambda\to u(y;\lambda)$ is analytic with poles
corresponding to values of $\lambda$ that are eigenvalues for specific
spectral equations.  It is known that two spectral equations are
necessary to reconstruct the potential $q(y)$ from spectral data
\cite{Borg-46,CCPR-SIAM97,GelLev-51,McLaughlin}. However, we will see
that our problem allows extraction of at most only one spectrum.
Nonetheless, if the ``potential'' $q(y)$ is known for half of the
entire domain, one eigenvalue spectrum suffices for the complete and
unique reconstruction of $q(y)$ \cite{CCPR-SIAM97,RS-IP92}.

Since $q(y)$ is known for $0 \leq y \leq h(x_b)$,  we require

\begin{equation}
h(x_b) = {y^{*} \over 2} \equiv {h(1) \over 2}
\label{HBCOND}
\end{equation}

\noindent in order for $q(y)$ to be completely reconstructed on $[0,y^{*}]$
Therefore, the original functions $\mu(x)$ and $\sigma(x)$ must at
least be known for $0\leq x \leq x_b$, where

\begin{equation}
x_b = h^{-1}\left(\dfrac{y^{*}}{2}\right) = h^{-1}\left({h(1) \over 2}\right).
\label{BCOND}
\end{equation}

%we can solve
%(\ref{eq:sch}) on $[0,x_b]$ using the measured data boundary condition 
%$u(y_{0};\lambda)= w(x_{0};\lambda)$. Thus, we deduce $u'(y_{b};\lambda)$ from
%the measured data $u(0;\lambda)$. 

\noindent The full problem on $[0,y^{*}]$ becomes

\begin{equation}
  \label{eq:schdata}
  \begin{array}{l}
  u''(y) - q(y) u(y)= \lambda u(y), \qquad 0<y<y^{*}\equiv h(1) \\
  u(y^{*})=\dfrac{1}{f(y^{*})}\neq 0, \quad 
   \displaystyle {\partial u(0;\lambda)\over \partial y}=0. \\
%\quad \displaystyle {\partial u(0;\lambda) \over \partial y}=0,\\
  u(y_{0};\lambda),\, \quad\mbox{ known.}
 \label{SL}
  \end{array}
\end{equation}

Now, consider the associated eigenvalue problem

\begin{equation}
  \label{eq:spect}
   \begin{array}{l}
  \phi_k''(y)-q(y)\phi_k(y) = -\lambda_k \phi_k(y),\quad 0< y< y^{*},
    \quad k\geq1  \\
\phi_{k}'(0) =\phi_{k}(y^{*})=0. 
  \label{EVP}
  \end{array}
\end{equation}

\noindent If the $\lambda_{k}$ in (\ref{EVP}) can be determined from
our measured data $u(y_{0};\lambda)$, then $q(y)$ on $[0,y^{*}]$ can
be determined upon using classic theories described in
\cite{Borg-46,CCPR-SIAM97,GelLev-51,HL,McLaughlin,RS-MC92}.  To find
the eigenvalues to (\ref{EVP}), as well as the transformed threshold
$y^{*}$, we decompose the Green's function for (\ref{SL}) in terms of
the eigenfunctions of (\ref{EVP}), and obtain

\begin{equation}
u(y;\lambda) = -\sum_{k}\left[ {\phi_{k}(y)\phi_{k}'(y^{*})
u(y^{*})\over \lambda + \lambda_{k}}\right].
\label{EE}
\end{equation}

\noindent First, consider the setting of {Theorem 1}.  Since
$u(y^{*}) = 1/f(y^{*})>0$ and $|\phi_{k}'(y^{*})|>0$ (since $\phi_{k}$
is a non-trivial solution to (\ref{EVP})),
$|u(y_{0};\lambda)|\to\infty$ as $\lambda \to -\lambda_k$, {\it
  provided} $\phi_{k}(y_{0})\neq 0$.  The latter constraint is not
satisfied for all choices of $y_0$.  Let us define
\begin{equation}
  \label{eq:lambday}
  \Lambda_y = \{ y\in (0,y^*), \mbox{ there exists } k\geq1 \mbox{ such that }
        \phi_k(y)=0\}.
\end{equation}
Since the number of zeros of solutions to the Sturm-Liouville problem
(\ref{EVP}) is finite, the set $\Lambda_y$ is countable and therefore
of Lebesgue measure $0$ on $(0,y^*)$. As long as our initial
measurement position $y_0$ is not part of the above countable subset
of values where the eigenfunctions of (\ref{EVP}) have a node, we can
in principle obtain one spectrum by analytically continuing the data
$u(y_{0};\lambda)$ in $\lambda$ and finding the poles. The constraint
on $y_0$ also amounts to saying that the eigenfunctions $\phi_k$ cannot
be restricted to being eigenvectors of the same problem posed on the
domain $(0,y_0)$ instead of $(0,y^*)$. In the primitive variables the
countable set of forbidden measurement points is thus defined as
\begin{equation}
  \label{eq:countset}
  \Lambda = h^{-1}(\Lambda_y) = \{ x\in(0,1), h(x) \in \Lambda_y\}.
\end{equation}

Now consider {Theorem 2}. We can extend $u(y;\lambda)$ by evenness on
$(-y^*,y^*)$ and then by periodicity on $\Rm$. The resulting function
is then continuous in $y$ and we deduce from (\ref{EE}) that
\begin{equation}
\pdr{u}{y}(y;\lambda) = -\sum_{k}\left[ {\phi'_{k}(y)\phi_{k}'(y^{*})
u(y^{*})\over \lambda + \lambda_{k}}\right].
\label{EEprime}
\end{equation}
Now since $\mu(x)$ and $\sigma(x)$ are known on $(0,x_0)$ then so is
$q(y)$ on $(0,y_0)$. Since $u(y_0,\lambda)$ is known, we can solve
(\ref{eq:schforu}) for $u(y)$ on $(0,y_0)$ and thus obtain the value
of $\frac{\partial u}{\partial y}(y_0;\lambda)$. Since {\em either}
$|\phi_{k}(y_{0})|>0$ or $|\phi'_{k}(y_{0})|>0$ (for otherwise
$\phi_k(y)\equiv0$), we deduce that
$|u(y_{0};\lambda)|+|\frac{\partial u}{\partial
  y}(y_0;\lambda)|\to\infty$ as $\lambda \to -\lambda_k$.  Under the
hypotheses of {Theorem 1} and {Theorem 2}, we thus find that one can
reconstruct the spectrum $\{\lambda_k\}_{k\geq1}$ in (\ref{eq:spect})
from the measurements.

Finally, we must determine $y^{*}$.  Asymptotic expansions of the
eigenvalues for large $k$, found using the WKB approximation
\cite{BO,CCPR-SIAM97,RS-MC92} have the form:

\begin{equation}
\lambda_k  = \Big[\Big(k-\frac12\Big)\frac{\pi}{y^*}\Big]^2
   + \dfrac{1}{y^*}\dint_0^{y^*} q(y)dy + a_k,
\label{eigenwkb}
\end{equation}
where the sequence $a_k\in l^2$, i.e.,  $\|a_k\|=(\sum_{k=1}^\infty
a_k^2)^{1/2}<\infty$.  Equation (\ref{eigenwkb}) implies that the
threshold value $y^{*}$ can be evaluated using
\begin{equation}
  \label{eq:recbeta}
  y^{*} = \lim\limits_{k\to\infty} {k\pi \over \sqrt{\lambda_{k}}}.
\end{equation}
Thus, the domain size is reconstructed. The classical theories found
in \cite{Borg-46,CCPR-SIAM97,GelLev-51,HL} can then be directly
applied to uniquely reconstruct the remaining, unknown part of $q(y)$.
Moreover, efficient numerical algorithms have been developed for
reconstructing $q(y)$ \cite{brown,GelLev-51,RS-MC92}, and ultimately
$\mu$ and $\sigma$ from the spectral data. However, the reconstruction
of these spectral data from measurements is based on analytic
continuation. This implies that the inverse problem is severely
ill-posed \cite{engl} and that noise in the data is quite strongly
amplified in the reconstruction; see section \ref{sec:impl}. Note that
a similar approach was used in \cite{Ramm-JMAA-01} in the context of
the one-dimensional heat equation.

%%%%%%%%%%%%%%%%%%%%%%%%%%%%%%%%%%%%%%%%%%%%%%%%%%%
\section{Reconstruction of drift or diffusion term}
\label{sec:recfinal}
%%%%%%%%%%%%%%%%%%%%%%%%%%%%%%%%%%%%%%%%%%%%%%%%%%%

We showed in the preceding section that $q(y)$ and $y^{*}$ are
uniquely determined from the measurements provided that one set of
hypotheses stated in {Theorems 1-3} (cf. Section 6) are verified.
Combining equations (\ref{eq:f}) and (\ref{eq:q}), we find

\begin{equation}
  \label{eq:riccati}
  \nu'(y)+\frac{\nu^2}{2} = 2q(y),\qquad 0 < y < y^{*}.
\end{equation}
Since both $\mu$ and $\sigma$ are known on $[0,x_b]$, 
\begin{equation}
  \label{eq:riccatibc}
   \nu(y_{b})= \left({\sqrt{2}\mu(x_b)\over \sigma(x_b)}
   -{\sigma'(x_b)\over \sqrt{2}}\right).
\end{equation}
%Upon using the explicit form for $f(0\leq y\leq y^{*})$ found from 
%solving (\ref{eq:f}) in the expression (\ref{eq:q}), 
The Riccati equation (\ref{eq:riccati}), along with the boundary condition 
(\ref{eq:riccatibc}) admits a unique solution for $\nu(y)$.

Even if $\nu(0\leq y\leq y^{*})$ is known uniquely, 
one cannot extract both $\mu(x)$ and $\sigma(x)$ independently.
However, if
either $\mu$ {\it or} $\sigma$ is known, then the other one can
uniquely be reconstructed from the knowledge of $\nu(y)$ on
$(0,y^{*})$. The simplest case occurs when only $\sigma$ is known, for
then, so is $h(x)$, and the reconstruction of $\mu$ follows from the
definition (\ref{eq:nu}) of $\nu$.  On the other hand, we can deduce
from (\ref{eq:h}) and (\ref{eq:nu}) that
\begin{equation}
 \left({\sigma'\over 2}-{\mu\over \sigma}\right)(x) =-{1\over  \sqrt{2}}
       \nu\Big(\int_0^x \frac{\sqrt2}{\sigma(x')}dx'\Big).
\label{eq:case2}
\end{equation}
Let us assume that $\mu/\sigma$ can be expressed as a known (smooth)
function of $\sigma$ and $x$.  This includes the cases when $\mu$ is
known or when $\mu/\sigma$ is known. Since $\sigma$ is {\it a priori}
known to be uniformly bounded from below, equation (\ref{eq:case2})
provides a first-order integro-differential equation that can then be
solved for $\sigma(x)$ on $(0,1)$. The reconstruction is uniquely
defined, concluding the proof of the main theorem.

%%%%%%%%%%%%%%%%%%%%%%%%%%%%%%%%%%%%%%%%%%
%%%%%%%%%%%%%%%%%%%%%%%%%%%%%%%%%%%%%%%%%%
%%%%%%%%%%%%%%%%%%%%%%%%%%%%%%%%%%%%%%%%%%
%%%%%%%%%%%%%%%%%%%%%%%%%%%%%%%%%%%%%%%%%%

%\paragraph{Remarks on the reconstruction.}

The reconstruction of $q(y)$ on $y\in (0,y^{*})$ clearly depends on
the choice of $q(y)$ on $[0,y_{b}]$ since it would change the spectrum
through Eq. (\ref{EVP}).  Therefore, a physically incorrect model of
$q$ on $[0,y_{b}]$, contaminates the reconstruction of $\mu(x),
\sigma(x)$ on $x\in (0,1)$, leading to physically incorrect, albeit
unique, results. To be more specific, consider the simple case
$\sigma^2\equiv2$, whence $h(x)=x$, $y^*=1$, and $\nu(x)=\mu(x)$.
Provided $|\mu|^2\ll|\mu'|$, Eq. \ref{eq:q} implies $q(x)\sim
\mu'(x)/2\ll1$. We then perform an asymptotic expansion of the
eigenfunctions $\phi_k=\phi_{k0}+\phi_{k1}$ and eigenvalues
$\lambda_k=\lambda_{k0}+\lambda_{k1}$, where $\phi_{k1}\ll1$ and
$\lambda_{k1}\ll1$ (see \cite{RS-IP92} for instance), and obtain

\begin{displaymath}
  \lambda_{k0} = \Big(\pi\Big(k-\dfrac12\Big)\Big)^2,\qquad\mbox{ and }
  \qquad \phi_{k0}(x) = \cos \pi\Big(k-\dfrac12\Big)x,
\end{displaymath}
and using the Fredholm alternative in the perturbation
\begin{displaymath}
   -\phi_{k1}'' + q \phi_{k0} = \lambda_{k1}\phi_{k0}+\lambda_{k0}\phi_{k1},
\end{displaymath}

\noindent we obtain

\begin{displaymath}
  \lambda_{k1} = \dfrac{\dint_0^1 q(x) \phi^2_{k0}(x) dx}
                {\dint_0^1 \phi^2_{k0}(x) dx}.
\end{displaymath}
Using the relation $\cos 2\theta=2\cos^2\theta-1$, we deduce from the 
above equation that the knowledge of $\lambda_{k1}$ allows us to obtain
\begin{displaymath}
  q_{2m-1} = \dfrac12\dint_0^1 q(x) \cos (2m-1)\pi x \,dx.
\end{displaymath}
Therefore the knowledge of one spectrum $\lambda_k$ gives us half of
the cosine Fourier transform of $q(x)$, hence of $\mu(x)$.  The other
half needs to be known, for instance by imposing that we know $q(x)$,
or equivalently $\mu(x)$, on $(0,1/2)$.

Although we have shown what parts of $\mu(x)$ and $\sigma(x)$ can be
uniquely reconstructed from a single first-exit time distribution, in
practice, finding the eigenvalues $\lambda_{k}$ from data to the
precision required for accurate reconstruction is difficult. 
%For such
%problems then, more practical algorithms must be devised, and is
%beyond the scope of this work.
 
%%%%%%%%%%%%%%%%%%%%%%%%%%%%%%%%%%%%%%%%%%%%%%%%%%%
\section{Reconstruction with multiple measurements}
\label{sec:multiple}
%%%%%%%%%%%%%%%%%%%%%%%%%%%%%%%%%%%%%%%%%%%%%%%%%%%

We have seen in the preceding section that either $\mu$ or $\sigma$
could be reconstructed from the knowledge of $q$ provided the other
one is known. In this section we obtain sufficient conditions on the
measurements so that $q(x)$ can uniquely be reconstructed on
$(\eps,1)$ for $\eps>0$.  We show that it is sufficient to measure
$u(y_m;\lambda)$ at a countable sequence of points $y_m$ with
accumulation point at $y=0$.  These points $y_m$ correspond to points
$x_m$ via the transform $x_m=h^{-1}(y_m)$. They are constructed as
follows.

We choose $y_0$ such that $y_0\geq y^*/2$ and then by induction the
points $y_m$ ($m\geq1$) such that $y_m\in (y_{m-1}/2,y_{m-1})$.  We
assume that $y_m\to0$ as $m\to\infty$. We denote by ${\cal Y}$ the set
of sequences $\{y_m\}_{m\in\Nm}=(y_0,y_1,\cdots)$ satisfying the above
constraints. We can then show the following result
\begin{theorem}
  \label{thm:3}
  Let us assume that $q\in L^2(0,1)$, that there is a small interval
  $[0,\eps]$ where $q(y)$ is constant and that there is a known index
  $M$ such that $y_M<\eps$. The parameters $\eps$ and $y_M$ need not
  be known.  Then the measurements at $u(y_m;\lambda)$, $1\leq m\leq
  M+1$, for $\{y_m\}\in{\cal Y}$ and $\lambda>0$ uniquely determine
  $q(y)$ on $(0,y^*)$ as well as the point $\{y_m\}_{m\in\Nm}$ and
  $y^*$.
\end{theorem}
\begin{proof}
  By hypothesis, we know $M$ such that $y_M<\eps$.  Consider the value
  of $u(y_M;\lambda)\neq 0$. If $u(y_M;\lambda)=0$, the condition at
  $y=0$ would imply that $u(y;\lambda)\equiv0$ on $(0,y_M)$ and, hence
  on $(y_M,y^*)$, since then $u(y_M;\lambda)=\frac{\partial
    u}{\partial y}(y_M;\lambda)=0$.  This however contradicts the fact
  that $u(y^*)=1$. Upon dividing the solution $u(y;\lambda)$ by
  $u(y_M;\lambda)\not=0$ we deduce that
  $v_M(y;\lambda)=u(y;\lambda)/u(y_M;\lambda)$ solves
\begin{equation}
  \label{eq:schdata2}
  \begin{array}{l}
  v_M''(y;\lambda) - q(y) v_M(y;\lambda)= \lambda v_M(y;\lambda), 
   \qquad 0<y<y_M \\
  v_M(y_M;\lambda)=1, \quad 
   \displaystyle {\partial v_M(0;\lambda)\over \partial y}=0.
  \end{array}
\end{equation}
We then know from the proof of Theorem \ref{thm:2} and the asymptotic
expansion (\ref{eigenwkb}) that the measurement of $v(y_{M+1})$
uniquely determines $y_M$ and $q(y)$ on $(0,y_M)$ since the latter is
constant.

The rest of the proof follows by induction. Let us assume that $q(y)$
is known on $(0,y_{m+1})$ for $m+1\leq M$. Then $u(y_m;\lambda)\not=0$
and $v_m(y;\lambda)=u(y;\lambda)/u(y_m;\lambda)$ satisfies the same
equation (\ref{eq:schdata2}) with $y_M$ replaced by $y_m$. Upon using
Theorem \ref{thm:2} we deduce that the knowledge of
$v_{m}(y_{m+1};\lambda)$ for $\lambda>0$ uniquely determines $y_m$ and
$q(x)$ on $(0,y_m)$. This allows us to reconstruct $y_0$ and $q(y)$ on
$(0,y_0)$ by induction. Applying Theorem \ref{thm:2} on $(0,y^*)$ one
last time concludes the proof of the theorem.
\end{proof}

The same type of proof could be used with other local {\it a priori}
information on the behavior of $q(y)$ in the vicinity of $y=0$.  For
instance, we can show that the reconstruction is feasible provided that
$q(x)$ is known on an arbitrarily small interval $(0,\eps)$ and
provided that there exists an admissible sequence
$(y_0,y_1,\cdots,y_M)$ of measurements with $y_M\leq\eps$.

In both cases the reconstruction of $q(y)$ and of the measurement
points $y_m$ allows us to obtain some information about the original
drift and diffusion terms $\mu(x)$ and $\sigma(x)$. We can for
instance prove the following result.

\begin{theorem}
  \label{thm:4}
  Assume that $\mu(x)$ and $\sigma(x)$ are known on $(0,\eta)$ for
  some $\eta>0$ and that $w(x_m;\lambda)$ is measured for $\lambda>0$
  at $x_m$, $1\leq m\leq M$, where the constraints on $\{x_m\}_m$ are
  that $x_M\leq\eta$ and $\{y_m\}_{m} = \{h(x_m)\}_m \in {\cal Y}$.
  If, in addition, either $\mu(x)$, $\sigma(x)$, or $\mu(x)/\sigma(x)$
  is known on $[0,1]$, then both $\mu(x)$ and $\sigma(x)$ are uniquely
  determined on $[0,1]$.
\end{theorem}
\begin{proof}
  The proof follows by induction. We define
  $y_M=h(x_M)\leq\eps=h(\eta)$.  Let us now define $f_M(y)$ as in
  (\ref{eq:f}) on $(0,y_{M-1})$ but with boundary condition
  $f_M(y_{M})=1$. This implies that
  $u_{M}(y_M;\lambda)=w(x_M;\lambda)$ where $u_M(y;\lambda)$ solves
  the Schr\"odinger equation
  \begin{equation}
  \label{eq:schforu2}
  \begin{array}{l}
   u_M''(y) - q(y) u_M(y) =\lambda u_M(y), \qquad 0 < y < y_{M-1}\\
  u_M(y_{M-1})=\dfrac{1}{f(y_{M-1})}>0, \quad 
  \displaystyle {\partial u_M(0;\lambda) \over \partial y} = 0.
  \end{array}
  \end{equation}
  We have seen in Theorem \ref{thm:3} that $q(y)$ can then be uniquely
  reconstructed on $(0,y_{M-1})$ since it is known on $(0,y_M)$.  We
  can now apply the rest of the proof of Theorem \ref{thm:2} on the
  interval $(0,x_{M-1})$. This allows us to reconstruct $\mu$ and
  $\sigma$ uniquely on $(0,x_{M-1})$.  Similarly the knowledge of
  $\mu$ and $\sigma$ on $(0,x_{m})$ allows us to reconstruct both
  coefficients uniquely on $(0,x_{m-1})$ and the same knowledge on
  $(0,x_0)$ allows reconstruction on $(0,1)$.  This concludes the
  proof of the theorem.
\end{proof}

%%%%%%%%%%%%%%%%%%%%%%%%%%%%%%%%%%%%%%%%%%%%%%%%%%%
\section{Implementation of the reconstruction}
\label{sec:impl}
%%%%%%%%%%%%%%%%%%%%%%%%%%%%%%%%%%%%%%%%%%%%%%%%%%%

Theorems \ref{thm:1}, \ref{thm:2}, and \ref{thm:4} provide an explicit
method to reconstruct the drift and diffusion coefficients $\mu$ and
$\sigma$ from the measurements. This method is based on three steps.
The first step consists of finding the poles of meromorphic functions
of the form $\lambda\to u(y;\lambda)$ at fixed $y$ from its knowledge
for $\lambda>0$. In the second step, the potential $q(y)$ is reconstructed
from the knowledge of the spectral data provided in step 1. Finally
$\mu(x)$ and $\sigma(x)$ are reconstructed from $q(y)$ by solving 
first-order ordinary differential equations. 

Step 3 is well-posed in the sense that noise in the potential $q(y)$
are at most linearly amplified in the reconstruction of $\mu(x)$ and
$\sigma(x)$. Step 2 is also well posed as the reconstruction of the
potential $q(y)$ from the spectra is stable, for instance in the sense
that an error of order $\eps$ on the $l^2$ norm of $a_k$ in
(\ref{eigenwkb}) implies an error also of order $\eps$ in the
reconstruction of $q(x)$ in the $L^2$ sense \cite{McLaughlin2,RS-MC92}.

Step 1, however, is severely ill-posed in the sense that order-$\eps$
errors in the measured generate errors greater than order
$\eps^{\alpha}$ (for all $\alpha > 0$) in the $l^2$ norm of the
coefficients $a_k$ in (\ref{eigenwkb}).  This is a similar result to
the reconstruction of diffusion coefficients from boundary
measurements, which is known to be severely ill-posed
\cite{Al-JDE90,isakov-98}.  We can show a H\"older-type stability
result in the reconstruction of each eigenvalue. This means that for
each eigenvalue $\lambda_k$, there exists $\alpha_k>0$ such that
errors in the data of order $\eps$ yield an error of order
$\eps^{\alpha_k}$ on the reconstruction.  Unfortunately, the prefactor
in front of $\eps^{\alpha_k}$ grows exponentially in $k$ so that at
most only approximately $|\log\eps|$ eigenvalues can be reconstructed
from oder-$\eps$ accurate measured data.  This behavior is a
consequence of the results on the meromorphic prolongation of
functions with poles, as shown by K.  Miller \cite{miller-70}. Since
we do not attack the numerical reconstruction here we briefly sketch
the result and how it may be used to derive the above stability
estimates.

Let us consider the open disk $D$ of radius $R$ and center $0$ in the
complex plane and a meromorphic function $f(\lambda)=u(y_0;\lambda)$
inside that disk with $N$ poles, which we assume lie on the segment
$(-R,0)$. We want to reconstruct these $N$ poles from the measurement
of the meromorphic function on the segment $\Gamma=(0,R)$. Let us
define by $w(z)$ the function that takes the value $0$ on $\partial
D$, the value $1$ on $\Gamma$, and is harmonic on $D\backslash\Gamma$.
One verifies that $0<w(z)<1$ on $D\backslash\Gamma$. Let us recast our
meromorphic function as
$f(\lambda)=F(\lambda)/\prod_{k=1}^N(\lambda-\lambda_k)$ where
$F(\lambda)$ is analytic on $D$. We assume that
\begin{displaymath}
  |F-Bh|<\eps|B| \quad \mbox{ on } \quad  \Gamma, \qquad\mbox{ and } \qquad
  |F|<E, \quad \mbox{ on } \quad \partial D.
\end{displaymath}
Here $h(\lambda)$ are our measurements. The above relation indicates
what we mean by error in the measurements and essentially says that
$f-h$ is bounded by $\eps$ in the supremum norm. Upon slightly
modifying the proof of Theorem 1 in \cite{miller-70}, we obtain that
it is possible to reconstruct the poles $\lambda_k$ of $f$ on $D$ with
an error of order
\begin{displaymath}
 |\delta\lambda_k|\leq
  \dfrac{C^N R^N(\eps R^N)^{w(\lambda_k)}E^{1-w(\lambda_k)}}
   {|W_k| \Big(\prod\limits_{l\not=k}|\lambda_l-\lambda_k|\Big)^2}.
   %\qquad W_k = u(y^*) \phi_k(y_0)\phi_k'(y^*),
\end{displaymath}
Here we have bounded $B$ by $R^N$ on $D$.  In the case of interest in
this paper, we have that
\begin{displaymath}
  R \sim N^2, \qquad \lambda_k \sim k^2, \qquad \mbox{\rm and } \qquad
   W_k = u(y^*) \phi_k(y_0)\phi_k'(y^*).
\end{displaymath}
The radius is chosen as roughly $1/2(\lambda_N+\lambda_{N+1})$.  At
the boundary $\partial D$ of the domain, we observe that $F$ is
bounded by (using Stirling's approximation)
\begin{displaymath}
  E\sim \prod\limits_{l\not=k}|\lambda_l-\lambda_k| \sim (N!)^2 
  \approx N^{2N+1}e^{-N}.
\end{displaymath}
Using these values we find that 
\begin{displaymath}
  |\delta\lambda_k|\leq C^N \eps^{w(\lambda_k)}
\end{displaymath}
for some positive constant $C$. The above constraint is sharp
\cite{miller-70}. This shows that the reconstruction of any eigenvalue
is H\"older continuous, where the order of continuity
$\alpha=w(\lambda_k)$. This is therefore a mildly ill-posed problem.

However, although the reconstruction of each eigenvalue is mildly
ill-posed, collectively the reconstruction of the whole spectrum is
severely ill-posed if one wants an approximation in the $l^2$ sense.
Indeed for a given noise level $\eps$, the above formula shows that we
can only reconstruct a number of eigenvalues that is proportional to
$|\log\eps|$, for the reconstructed information becomes useless as
$|\delta\lambda_k|\geq1$ and the corresponding coefficients $a_k$ in
(\ref{eigenwkb}) are not reconstructed at all. Now assuming that
$w(\lambda_k)=1$, which is the best we can expect, we observe that
$C^N\eps=1$ for $N\sim|\ln\eps|$. This shows that the reconstruction
of the potential $q(y)$ is severely ill-posed. Indeed let us assume
that the asymptotic coefficient of $q$ in (\ref{eigenwkb}) are of the
form $a_k\sim k^{-M}$ for some $M$. The $l^2$ error on the Fourier
coefficients obtained by truncating all coefficients of index greater
than $\vert\ln\eps\vert$ is then of order $N^{-M+1/2}\sim
|\log\eps|^{-M+1/2}\gg\eps^\alpha$ for all $\alpha>0$.

%%%%%%%%%%%%%%%%%%%%%%%%%%%%%%%%%%%%%%%%%%%%%%%%%%%
\section{Summary and Conclusions}
\label{sec:summary}
%%%%%%%%%%%%%%%%%%%%%%%%%%%%%%%%%%%%%%%%%%%%%%%%%%%

We have shown that if $\mu$ and $\sigma$ are known in say, $x\in
[0,x_b]$, with $x_b=h^{-1}(y^{*}/2)$ (Eq. \ref{BCOND}), one of them
can be reconstructed from a single first-exit time distribution
measurement at $x_{0}\not\in\Lambda$.  The procedure is formally
carried out by transforming the backward Kolmogorov equation for the
first-exit time distribution $w(x;\lambda)$ into a Schr\"{o}dinger
equation in $u(y;\lambda)$ with a potential $q(y)$ that is a
functional of $\mu(x)$ and $\sigma(x)$.  If this potential is known in
the domain $0 \leq y \leq y_{b}\equiv h(x_b)$, $q(y)$ can be
reconstructed from the single eigenvalue spectrum obtained by
analytically continuing the solution $u(y;\lambda)$ (in $\lambda$) and
finding the poles. It is this pole-finding procedure that is severely
ill-posed. The asymptotic limit of the eigenvalues also determine the
extent of the transformed domain $y^{*}$.  From the reconstructed
$q(y)$, one can reconstruct $\mu(x)$ or $\sigma(x)$, only if the other
(or $\mu(x)/\sigma(x)$) is also known.  Finally, we have shown that
the measurement process and the determination of spectra can be
repeated a sufficient number of times to obtain $q(y)$ ({Theorem 3})
and $\mu$ or $\sigma$ ({Theorem 4}) on the entire domain provided
$q(y)$ and $\sigma$ or $\mu$ is initially known on a small interval
$(0,\varepsilon)$.

Our analysis in section \ref{sec:impl} shows that the proposed
inversion is severely ill-posed even if the hypotheses of the
aforementioned theorems are satisfied. However the reconstruction of
partial spectral data is mildly ill-posed so we believe the proposed
method can be implemented. Let us conclude with the following remark.
Our method is based on reconstructing the eigenvalues $\lambda_k$ from
the knowledge of $u(x_0;\lambda)$. The latter may however provide
additional information. For instance we observe from (\ref{EE}) that
the residuals $\phi_{k}(y)\phi_{k}'(y^{*})u(y^{*})$ can also be
reconstructed from the measured data. It is known that the measurement
of one spectrum $\{\lambda_k\}$ may be sufficient to uniquely
reconstruct $q(y)$ on the whole interval $(0,y^*)$ provided that
additional normalization conditions be also measured on the
eigenfunctions \cite{McLaughlin,RS-MC92}. It is not clear whether
additional information on $\phi_{k}(y)\phi_{k}'(y^{*})u(y^{*})$ may be
useful. Preliminary numerical simulations seem to show that
reconstructions from a single measurement only allows us to
reconstruct part of the potential.

%%%%%%%%%%%%%%%%%%%%%%%
\section*{Acknowledgments}
This work was supported by the National Science Foundation (US)
through grants DMS-0239097 (GB) and DMS-0206733 (TC) and by the Office
of Naval Research through grant N00014-02-1-0089 (GB). GB also
acknowledges support from an Alfred P. Sloan Fellowship.

%%%%%%%%%%%%%%%%%

%%%
%%%%%%%%%%%%%%%%%%%%%%%%
%
%%
%%%%%%%%%
\end{document}